\documentclass[epj]{svjour}
\usepackage{latexsym}
\usepackage{graphics}
\usepackage[latin1]{inputenc}
\begin{document}
\renewcommand{\textfraction}{0.00000000001}
\renewcommand{\floatpagefraction}{1.0}

\title{Double pion photoproduction off $^{40}$Ca}
\author{F.~Bloch\inst{1},
J.~Ahrens\inst{2},
J.R.M.~Annand\inst{3},
R.~Beck\inst{2},
L.S.~Fog\inst{3},
D.~Hornidge\inst{2},
S.~Janssen\inst{4},
M.~Kotulla\inst{1}
\thanks{\emph{Present address:}	II. Physikalisches Institut, 
              Universit\"at Giessen, D-35392 Giessen, Germany},
B.~Krusche\inst{1},
J.C.~McGeorge\inst{3},
I.J.D.~MacGregor\inst{3},
J.G.~Messchendorp\inst{4}
\thanks{\emph{Present address:}	Kernfysisch Versneller Institut (KVI), 
              Groningen, The Netherlands},
V.~Metag\inst{4},
R.~Novotny\inst{4},
R.O.~Owens\inst{3},
M.~Pfeiffer\inst{4},
M.~Rost\inst{2},
R.~Sanderson\inst{3},
S.~Schadmand\inst{4}
\thanks{\emph{Present address:}	Institut f\"ur Kernphysik, 
              Forschungszentrum J\"ulich, D-52425 J\"ulich, Germany},
A.~Thomas\inst{2},
D.P.~Watts\inst{3}
}                     
\offprints{Bernd.Krusche@unibas.ch}          %
\institute{Department of Physics and Astronomy, University of Basel,
           Ch-4056 Basel, Switzerland \and
	   Institut f\"ur Kernphysik, Johannes-Gutenberg-Universit\"at Mainz,
           D-55099 Mainz, Germany \and
	   Department of Physics and Astronomy, University of Glasgow, Glasgow
	   G12 8QQ, UK \and 
	   II. Physikalisches Institut, Universit\"at Giessen, D-35392
           Giessen, Germany 
           }
\date{Received: date / Revised version: date}
%
\authorrunning{F. Bloch et al.}
\titlerunning{Double pion photoproduction off $^{40}$Ca}

\abstract{The photoproduction of $\pi^0\pi^0$ and $\pi^0\pi^{\pm}$
pairs off $^{40}$Ca has been investigated with the TAPS detector
using the Glasgow photon tagging spectrometer at the Mainz MAMI accelerator.
Data have been taken for incident photon energies in the energy range from 
200 - 820 MeV. Total cross sections have been extracted from threshold up to the
maximum photon energy and invariant mass distributions of the pion pairs 
have been obtained for
incident photon energies between 400 - 500 MeV and 500 - 550 MeV. The double 
$\pi^0$ invariant mass distributions show some relative enhancement with 
respect to the mixed charge channel at small invariant masses. 
The effects are smaller than previously observed for lead 
nuclei and the distributions do not significantly deviate from carbon data. 
The data are in good agreement with the results of recent calculations in 
the framework of the BUU model, with careful treatment of final state 
interaction effects but without an explicit in-medium modification of scalar, 
iso-scalar pion pairs. This means that for Ca most of the experimentally 
observed effect can be explained by final state interactions. Only at low incident 
photon energies there is still a small low mass enhancement of the double 
$\pi^0$ data over the BUU results.
} 
\maketitle

\section{Introduction}
\label{sec:1}
In-medium properties of hadrons are a hotly debated topic since they are closely
related to the properties of low-energy, non-perturbative QCD. QCD at high
energies or short scales ($r < 0.1$fm) is well explored by perturbative methods.
However, at larger distances the perturbative picture breaks down and at scales
between 0.1 fm and 1 fm the full complexity of the interaction manifests itself
in the many body structure of hadrons composed of valence quarks, sea quarks,
and gluons. Chiral symmetry is at the very heart of the theory. In the limit of
vanishing current quark masses the QCD Lagrangian is invariant under chiral
rotations, right and left handedness of quarks is conserved and right and left
handed fields can be treated independently. The explicit breaking of the
symmetry due to finite u,d current quark masses (5-15 MeV) is small. However,
it is well known that spontaneous breaking occurs since the ground state,
the QCD vacuum, has only part of the symmetry. This is connected to a non-zero
expectation value of scalar $q\bar{q}$ pairs in the vacuum, the chiral
condensate. The symmetry breaking is clearly reflected in the hadron spectrum.
Without it, hadrons should appear as mass degenerate parity doublets, which is
neither true for baryons nor for mesons. The first $J^{\pi}$=1/2$^{-}$ excited
state in the baryon spectrum, the S$_{11}$(1535) lies much above the 
$J^{\pi}$=1/2$^{+}$ nucleon ground state. The lowest lying $J^{\pi}$=1$^{-}$
vector meson, the $\rho$, has a smaller mass than the $J^{\pi}$=1$^{+}$
$a_1$, and the $J^{\pi}$=0$^{-}$ pion (the Goldstone boson of
chiral symmetry), is much lighter than its chiral partner, the $J^{\pi}$=0$^{+}$ 
$\sigma$-meson. 

However, model calculations (see e.g. \cite{Lutz_92}) indicate a significant
temperature and density dependence of the chiral condensate which is
connected with a partial restoration of chiral symmetry at high temperature
and/or large densities. An observable consequence is a density and temperature 
dependence of hadron masses. These effects will be most 
pronounced for the high temperature/density conditions probed in heavy ion 
reactions, but they should be already significant at zero temperature and 
normal nuclear density as probed by pion and photon beams. In-medium
modifications of vector mesons have been searched for via the
spectroscopy of di-lepton pairs emitted in heavy ion reactions by the CERES
experiment \cite{Agakichiev_95,Adamova_03} and more recently the NA60 
collaboration \cite{Damjanovic_05}, which reported a strong broadening of the
in-medium spectral function of the $\rho$-meson. A signature for in-medium
modifications of $\rho$ and $\omega$ mesons was also reported from the E325 
experiment in 12 GeV p+A reactions at KEK \cite{Naruki_06}.  An in-medium mass 
shift of the $\omega$ meson has been found in photon induced $\omega$-production
from heavy nuclei \cite{Trnka_05}. Due to their coupling to vector mesons, 
nucleon resonances should also be modified in the nuclear medium.
Self-consistent calculations of the respective spectral functions 
\cite{Post_04} predict in particular significant effects for the 
D$_{13}$(1520) resonance, which have also been searched for in photon 
induced meson production reactions 
\cite{Krusche_01,Krusche_04,Krusche_04a,Krusche_05}, however so far
without conclusive results. As discussed in detail by Lehr and Mosel
\cite{Lehr_01}, part of the problem is, that due to the averaging over the 
nuclear volume, a possible broadening of the resonance can be only observed 
via decay channels which cause the broadening, but not in other exclusive 
reactions. 
 
A particularly interesting case is the mass-split between the $J^{\pi}$=0$^{-}$ 
pion and the $J^{\pi}$=0$^{+}$ $\sigma$-meson. The naive assumption that the 
two masses should become degenerate in the chiral limit is supported by model
calculations. A typical result is the density dependence of the mass calculated
in the Nambu-Jona-Lasino model by Bernard, Meißner and Zahed 
\cite{Bernard_87}. The $\sigma$-mass drops significantly as function of nuclear 
density. Compared to the vacuum  the predicted effect is already large  for 
normal nuclear matter density $\rho_0$ at which the pion mass stays still stable. 
The nature of the $\sigma$ meson itself has been controversely discussed in 
the literature. 
The review of particle properties \cite{Yao_06} lists as $\sigma$ 
the $f_{0}$(600) with a mass range from 400 - 1200 MeV and a full width 
between 600 MeV and 1000 MeV. Recently, Caprini, Colangelo and Leutwyler
\cite{Caprini_06} have derived precise predictions for mass and width from 
dispersion relations. In some approaches it is treated as a pure 
$q\bar{q}$ (quasi)bound state \cite{Bernard_87,Hatsuda_99,Aouissat_00}, 
in other approaches as a correlated $\pi\pi$ pair in a $I=0$, $J^{\pi}=0^+$ 
state \cite{Chiang_98,Roca_02,Chanfray_06}. But in any case a strong coupling to  
scalar, iso-scalar pion pairs is predicted. As a consequence, the different
model approaches agree in so far, as they predict a significant in-medium
modification of the invariant mass distribution of such pion pairs. This is
either due to the in-medium spectral function of the $\sigma$ meson
\cite{Hatsuda_99} or the in-medium modification of the pion-pion interaction
\cite{Roca_02} due to the coupling to nucleon - hole, $\Delta$ - hole and
$N^{\star}$ - hole states. In any case the predicted net effect is a downward
shift of the strength in the invariant mass distributions of scalar, 
iso-scalar pion pairs in nuclear matter.

First experimental evidence for this effect has been reported by the CHAOS
collaboration from the measurement of pion induced double pion production
reactions \cite{Bonutti_96,Bonutti_99,Bonutti_00,Camerini_04,Grion_05}.
The main finding was a build-up of strength with rising mass number at low
invariant masses for the $\pi^+\pi^-$ final state, but not for the 
$\pi^+\pi^+$ channel where the $\sigma$ as neutral particle cannot contribute. 
A similar effect was found by the Crystal Ball collaboration at BNL. They observed a 
low-mass enhancement of strength for heavy nuclei in the 
$\pi^- A\rightarrow A\pi^0\pi^0$ reaction \cite{Starostin_00}. 

Pion induced reactions have the disadvantage that the signal is diluted by
final (FSI) and initial (ISI) state interaction of the pions, so that
effectively only the low-density surface zone of the nuclei is probed. 
Pions can be produced in the entire volume of the nuclei by photon induced
reactions, although their FSI also suppresses the contributions 
from the deep interior of the nuclei. The FSI can be minimized by the use
of low incident photon energies, giving rise to low-energy pions which have
much larger mean free paths than pions that can excite the $\Delta$-resonance
\cite{Krusche_04}. Photoproduction of the different charge
states of pions from the free proton and the quasifree neutron has been
previously studied in detail with the 
DAPHNE \cite{Braghieri_95,Zabrodin_97,Zabrodin_99,Ahrens_03,Ahrens_05}
and TAPS detectors
\cite{Haerter_97,Krusche_99,Wolf_00,Kleber_00,Langgaertner_01,Kotulla_04}
at MAMI in Mainz from threshold to the second resonance region 
and for the $\pi^0\pi^0$ channel also at higher incident photon energies at 
GRAAL in Grenoble \cite{Assafiri_03} (see \cite{Krusche_03} for an overview). 

First results from a measurement of double $\pi^0$ and $\pi^0\pi^{\pm}$
photoproduction off carbon and lead have been reported in 
\cite{Messchendorp_02}. A shift of the strength to lower invariant masses was found
for the heavier nucleus for the $\pi^0\pi^0$ channel but not for the mixed charge channel.
In the present paper, experimental details and the results of an
additional measurement of double pion 
photoproduction off calcium nuclei are presented  and compared to model 
calculations. 
 
\section{Experimental setup}
\label{sec:2}

The experimental setup was identical to the measurement of double pion
photoproduction off carbon and lead nuclei reported in \cite{Messchendorp_02}.
Some further experimental details have been discussed in \cite{Krusche_04}.

\begin{figure}[bh]
\resizebox{0.49\textwidth}{!}{%
  \includegraphics{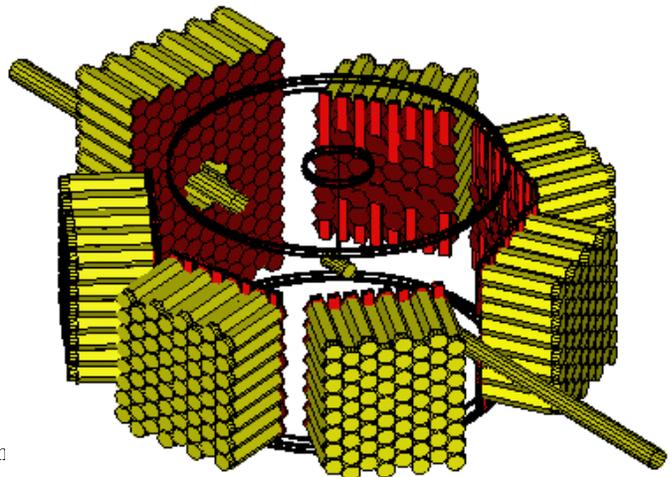}
}
\caption{Setup of the TAPS detector at the Mainz MAMI accelerator.
The beam entered the target chamber from the lower right edge. 
}
\label{fig:taps}       
\end{figure}

Quasi-monochromatic photons were produced with the Glasgow tagged photon
facility \cite{Anthony_91} at the Mainz MAMI accelerator \cite{Walcher_90}. 
The experiment covered the photon energy range from 200 - 800 MeV with an
energy resolution of approximately 2 MeV per tagger channel. 
The target was a 10 mm long cylinder of calcium with 50 mm diameter and the
beam-spot size at the target position was approximately 30 mm.
The reaction products from the target were detected with the electromagnetic
calorimeter TAPS \cite{Novotny_91,Gabler_94}. This detector consists
of 510 hexagonally shaped BaF$_2$ crystals of 25 cm length with an inner 
diameter of 5.9 cm. As shown in fig. \ref{fig:taps} they were arranged
in six blocks of 64 modules and a larger forward wall of 138 modules.
The blocks were arranged in one plane around the target at a distance of
55 cm from the target center and at polar angles of $\pm 54^o$, $\pm 103^o$, 
and $\pm 153^o$, while the forward wall was placed 60 cm away from the target 
center at 0$^o$ and the photon beam passed through a hole in the center of the
forward wall. Each detector module is equipped with an individual plastic veto
detector, read out by a separate photomultiplier. The setup covered approximately
 37 \% of the full solid angle.

\section{Data analysis}
\label{sec:3}

\begin{figure}[h]
\resizebox{0.50\textwidth}{!}{%
  \includegraphics{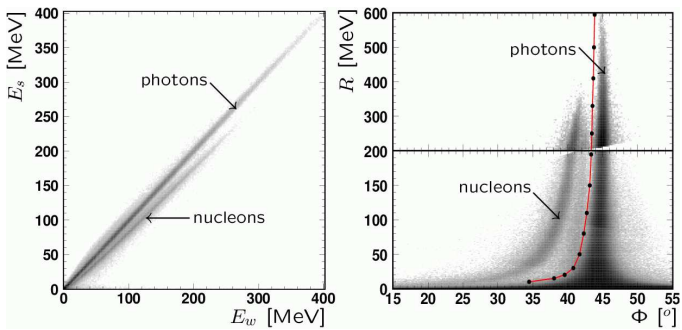}
}
\caption{Pulse shape analysis. Left hand side: plot of $E_{s}$ versus $E_{w}$.
Right hand side: plot in polar coordinates. 
}
\label{fig:psa}       
\end{figure}

The first step of the analysis for both final states - $\pi^0\pi^0$ and
$\pi^0\pi^{\pm}$ - requires the identification of photons. This was done with 
the help of the veto detectors, a time-of-flight analysis (time resolution:
550 ps FWHM), and a pulse shape analysis of the signals from the BaF$_2$ 
scintillator. The latter is based on the two different components of the
scintillation light from BaF$_2$ crystal, which have very different decay times.
The relative intensity of the fast component depends on the ionization density
produced by the incident particle, in particular it is suppressed for nucleons 
in comparison to photons. This feature was exploited by integrating the output 
signals over a short gate (40 ns) and a long gate (2 $\mu$s). 
Both signals $E_{s}$ (short gate) and $E_{w}$ (wide gate) were calibrated for 
photons so that in an $E_{s}$-versus-$E_{w}$ plot photons are lying on the 45$^o$ 
line. This is shown in the left hand side of fig. \ref{fig:psa}.
For the discrimination between photons and nucleons a representation in polar 
coordinates (see right hand side of fig. \ref{fig:psa}) is more efficient. 
The respective coordinates are the radius $R=\sqrt{E_s^2+E_w^2}$ and the 
angle $\Phi=arctan(E_s/E_w)$. The combination of the different discrimination 
methods provides a clean sample of photons. 

\begin{figure}[h]
\resizebox{0.50\textwidth}{!}{%
  \includegraphics{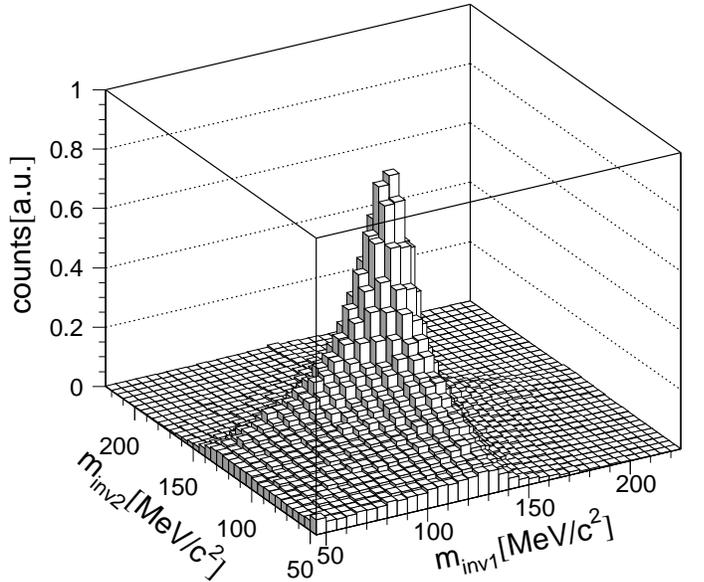}
}
\caption{Invariant mass of photon pair 1 versus invariant mass of photon pair
2 for events with four detected photons. 
}
\label{fig:2pi0minv}       
\end{figure}

For the investigation of the $\pi^0\pi^0$ final state events with four photons were
selected. The invariant masses of all possible combinations of the four photons
into two pairs were calculated and the `best' combination was chosen via a $\chi^2$
analysis which minimizes 
\begin{equation}
\chi^2 = \frac{(m_{\gamma_i,\gamma_j}-m_{\pi})^2}{\Delta m_{\gamma_i,\gamma_j}} +
\frac{(m_{\gamma_k,\gamma_l}-m_{\pi})^2}{\Delta m_{\gamma_k,\gamma_l}}
\end{equation}
with $1\leq i\neq j\neq k\neq l \leq 4$, where $m_{\pi}$ is the pion mass,
the $m_{\gamma ,\gamma}$ are the invariant masses of the photon pairs, and
the $\Delta m_{\gamma ,\gamma}$ are the associated uncertainties. A two-dimensional 
representation of the invariant masses of these pion pairs is shown in 
fig. \ref{fig:2pi0minv}. Events with both invariant masses in the range 
100 - 150 MeV were selected. 

At incident photon energies below the $\eta$ production threshold this analysis
results in a practically background-free data sample for the $\pi^0\pi^0$ final 
state. Above the $\eta$ threshold background arises from 
$\eta\rightarrow\pi^0\pi^0\pi^0$ decays, where the decay photons from one $\pi^0$
escape detection due to the limited solid angle of the calorimeter. This background
can be partly suppressed by a missing mass analysis. It is assumed that the reaction
occurs quasi-free off one participant nucleon. The initial momentum of this bound
nucleon is neglected. The mass of the missing particle - the participant nucleon - 
is then determined by the energy of the incident photon $E_b$ and the four-vectors 
of the four decay photons apart from the effects of Fermi motion, which broaden 
the missing mass distributions. 
\begin{figure}[th]
\resizebox{0.50\textwidth}{!}{%
  \includegraphics{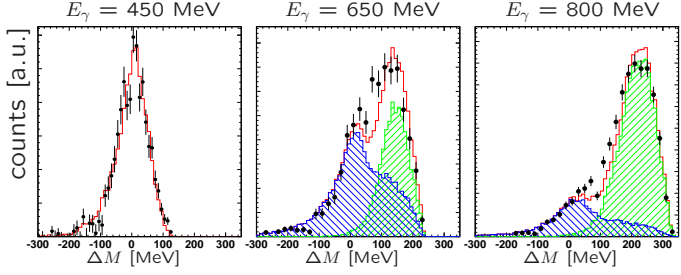}
}
\caption{Missing mass spectra for double $\pi^0$ photoproduction from
Ca. Left side: incident photon energy range 400 - 500 MeV (below $\eta$ threshold),
middle and right hand side above $\eta$ threshold$: E_{\gamma}$ around 650 and 800 
MeV, respectively. Closed circles: experiment, shaded histograms: simulation of 
$2\pi^0$ photoproduction (around zero) and $\eta$ photoproduction, solid histograms: 
sum of both. 
}
\label{fig:2pi0mis}       
\end{figure}
The missing mass $\Delta M$
is given by:
\begin{equation}
\Delta M = \sqrt{(E_b+m_N-\sum_{i=1}^{4}{E_{\gamma_i}})^2
-(\vec{P_b}-\sum_{i=1}^{4}{\vec{P_{\gamma_i}}})^2}- m_N.
\label{eq:mismas}
\end{equation}
where $E_{\gamma_i}$ and $\vec{P_{\gamma_i}}$ are energies and momenta of the decay
photons and $m_N$ is the nucleon mass. The resulting distributions are shown in fig.
\ref{fig:2pi0mis} for energies below and above the $\eta$ production threshold.
They are compared to Monte Carlo simulations which take into account Fermi motion,
a simple approximation of final state interaction (FSI) of the mesons, and the
detector acceptance and detection efficiency. The simulation of the detector response
was based on the GEANT3 package \cite{Brun_86}. At incident photon energies below
the $\eta$ threshold the data agree very well with the results of the simulation and
no indication of background is visible. The background from $\eta$ decays
is prominent at higher incident photon energies, but the data can still be reproduced
by a superposition of simulated double $\pi^o$ and $\eta\rightarrow\pi^0\pi^0\pi^0$
events. Due to the large broadening of the missing mass structures by Fermi motion and
FSI (responsible for the shoulders of the double $\pi$ distributions) a clear 
separation of the two reaction channels is not possible. However, the cross section 
for $\eta$ photoproduction is simultaneously measured via their 
$\eta\rightarrow\gamma\gamma$ decay. With this cross section and the known decay 
branching ratios it is possible to simulate the contribution of 
$\eta\rightarrow\pi^0\pi^0\pi^0$ decays to the analysis of double $\pi^0$ decays.
If this is done without any cut on missing mass, the result becomes independent
of the simulated shapes of the distributions, but the statistical quality suffers
since at high incident photon energies the $\eta$ background contribution is large.
Therefore as a compromise only events with missing mass between -100 and 100 MeV
were selected, so that the background contribution from the $\eta$ decays did not
exceed 20 \%.

\begin{figure}[t]
\resizebox{0.49\textwidth}{!}{%
  \includegraphics{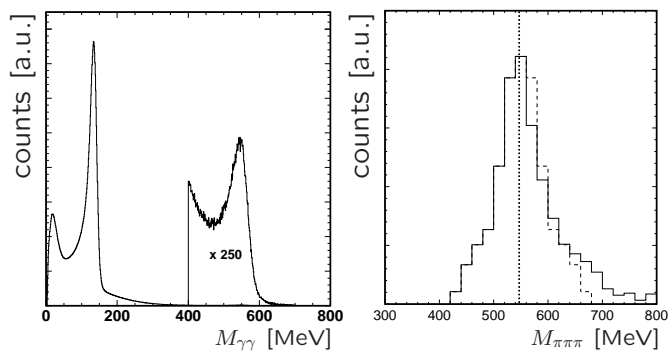}
}
\caption{Left hand side: invariant mass spectrum of photon pairs. Right hand side:
invariant mass spectrum of $\pi^0\pi^{\pm}\pi^{\pm}$ events. The dotted line indicates
the nominal position of the $\eta$ peak. The dashed histogram is the mirror image of
the left hand side of the peak (see text). 
}
\label{fig:2ginv}       
\end{figure}

The analysis of the $\pi^0\pi^{\pm}$ final state is more difficult. In the first 
step events with two photons and the invariant mass of the photon pair
(see fig. \ref{fig:2ginv}, left hand side) between 100 and 150 MeV were selected.
In the second step an additional particle was required. 
\begin{figure}[ht]
\resizebox{0.50\textwidth}{!}{%
  \includegraphics{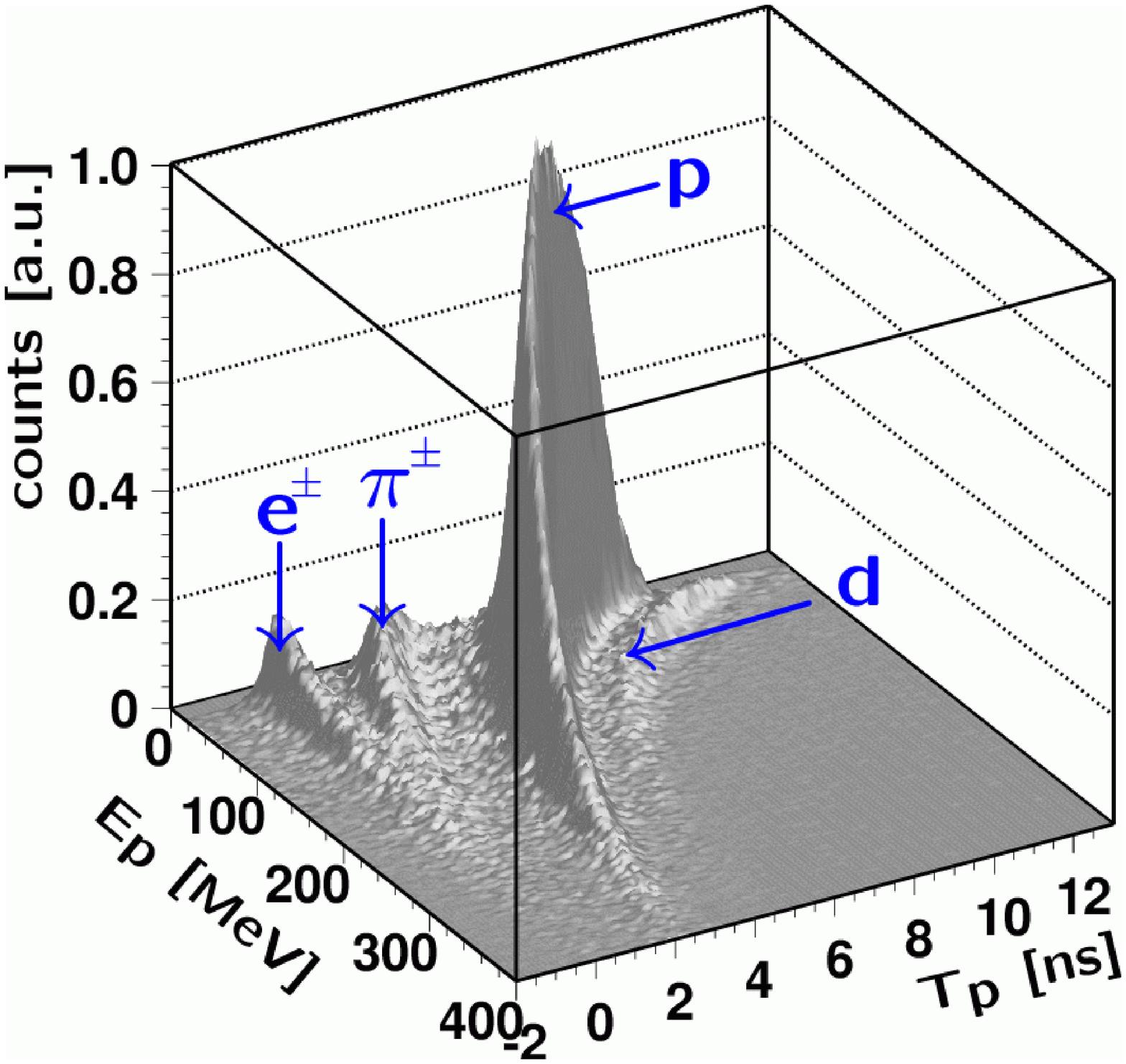}
}
\end{figure}
\begin{figure}[h]
\resizebox{0.50\textwidth}{!}{%
  \includegraphics{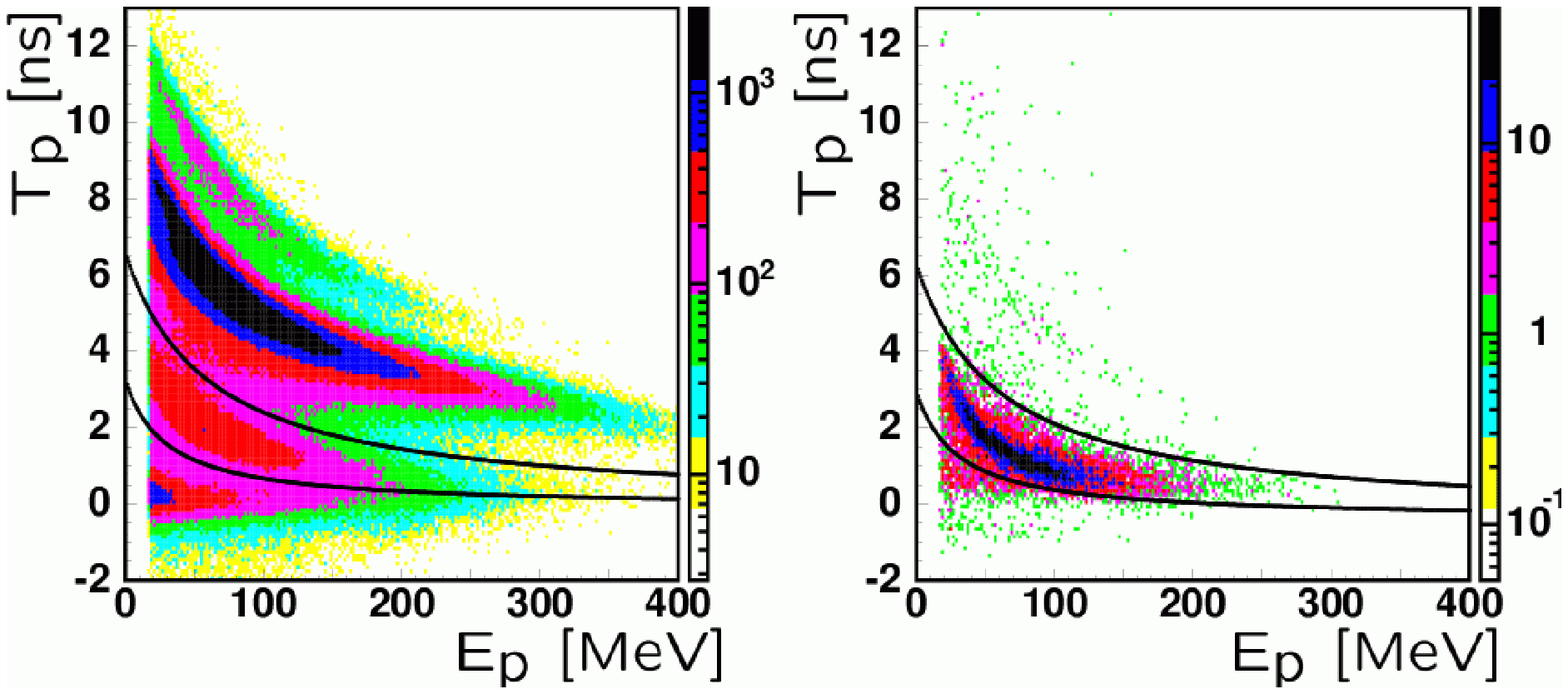}
}
\caption{Time-of-flight versus energy. The upper part shows a three-dimensional
representation of the TOF-versus-E spectrum. The bands for electrons, charged pions,
protons, and even deuterons are clearly visible. The bottom part shows a projection.
Left hand side: data, right hand side MC simulation for the $\pi^0\pi^{\pm}$ reaction
(only the charged pions were tracked by GEANT). The lines indicate the applied cuts.
}
\label{fig:tof_ener}       
\end{figure}
Candidates for 
charged pions were then selected by a time-of-flight versus energy analysis. 
The corresponding distributions of deposited energy $E_{p}$ versus time-of-flight 
$T_{p}$ are summarized in fig, \ref{fig:tof_ener}. They are
compared to the results of an MC simulation of $\pi^0\pi^{\pm}$ production. The shape
of the pion bands in the data and the simulation are in good agreement. It should be
noted, that $\pi^+$ and $\pi^-$ behave identically in the TOF-versus-E spectra.
This is not trivial since the $\pi^-$ get absorbed by nuclei once they have deposited
their kinetic energy, while the $\pi^+$ decay via the chain 
$\pi^+\rightarrow\mu^+\nu_{\mu}\rightarrow e^+\nu_e\nu_{\mu}$ after they have been
stopped. However, the kinetic energy of the decay muons is small and their lifetime
(2.2$\times 10^{-6}$ s) is much longer than the electronic gates so that the
energy deposition from the decay positrons is not measured. In the MC simulations
particles were only tracked for times corresponding to the hardware gate lengths
used in the experiment.
Since the time-of-flight is more precisely known than the deposited energy of the
pions, their kinetic energies were re-calculated from the time-of-flight.
The region accepted for pion candidates in TOF-versus-E is indicated in fig.
\ref{fig:tof_ener}. In order to suppress background from 
$\eta\rightarrow\pi^0\pi^+\pi^-$ 
decays it was furthermore required in the event selection, that no
second particle fulfilled the charged pion condition.   

The pion band in the TOF-versus-E analysis is not background free 
(see fig. \ref{fig:tof_ener}). There is some chance that protons leak into it.
This happens for example for high energy protons which punch through the BaF$_2$
modules and thus deposit too little energy or for protons that scatter on their way 
to the detector (e.g. in the target chamber). 
These misidentified protons are in particular a problem at low incident photon 
energies, where the cross section for the potential background final state $p\pi^0$ 
from single $\pi^0$ production via the $\Delta$ resonance is larger than the 
$\pi^0\pi^{\pm}$ cross section by more than an order of magnitude.
A suppression of this background is only possible via the reaction kinematics.
However, one must take into account that protons, which are misidentified as pions,
are assigned wrong kinetic energies since the energies of charged particles are 
based on the time-of-flight measurement using the mass of the particle. 
Therefore, only constraints based on the kinematics of the $\pi^o$ mesons can be used.
At very forward angles this background was so large that the forward wall
of the detector setup was not used for the charged pion analysis, which means 
that charged pions were only identified for laboratory polar angles larger than 
22$^o$.
\begin{figure}[h]
\resizebox{0.50\textwidth}{!}{%
    \includegraphics{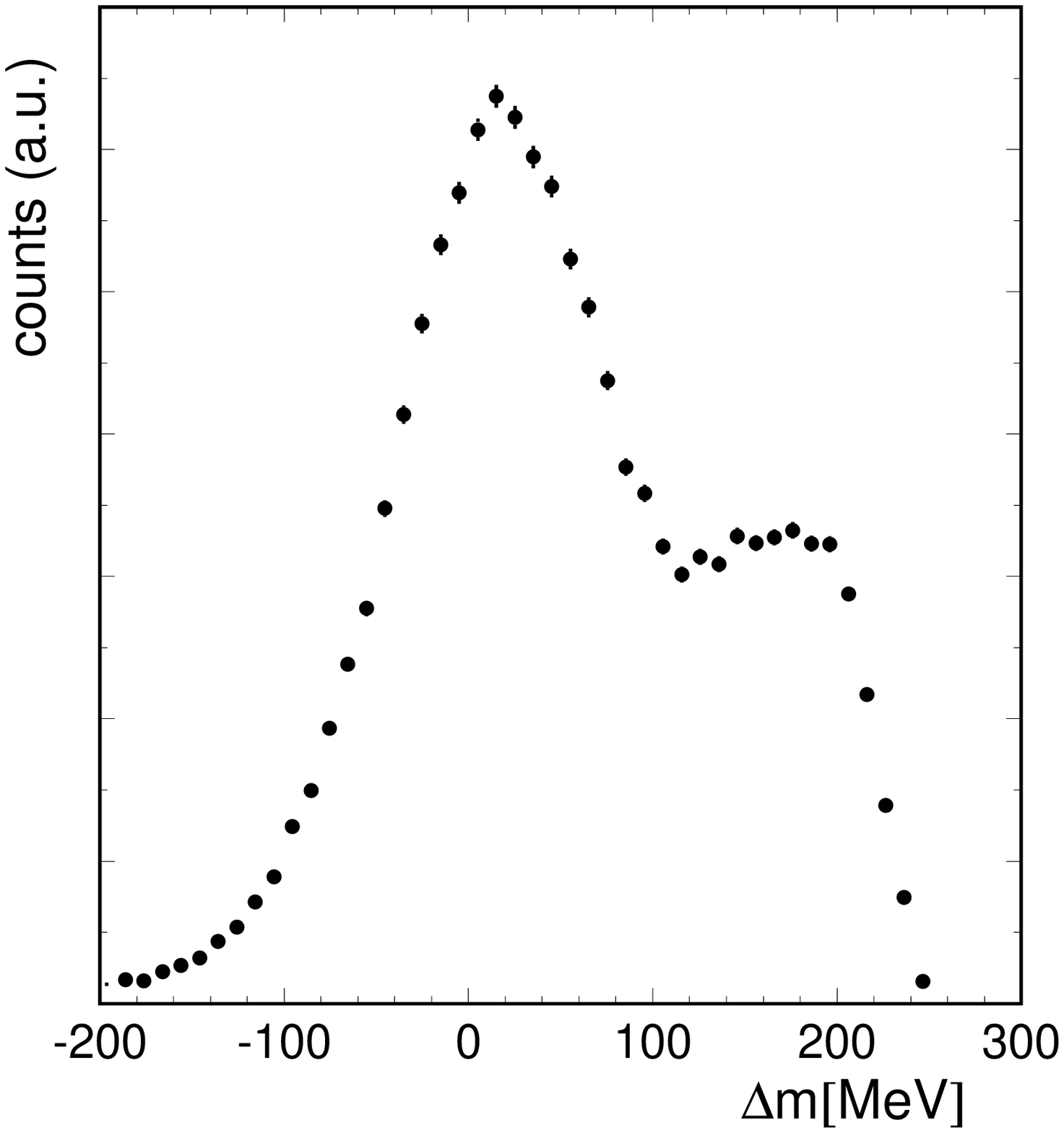}
    \includegraphics{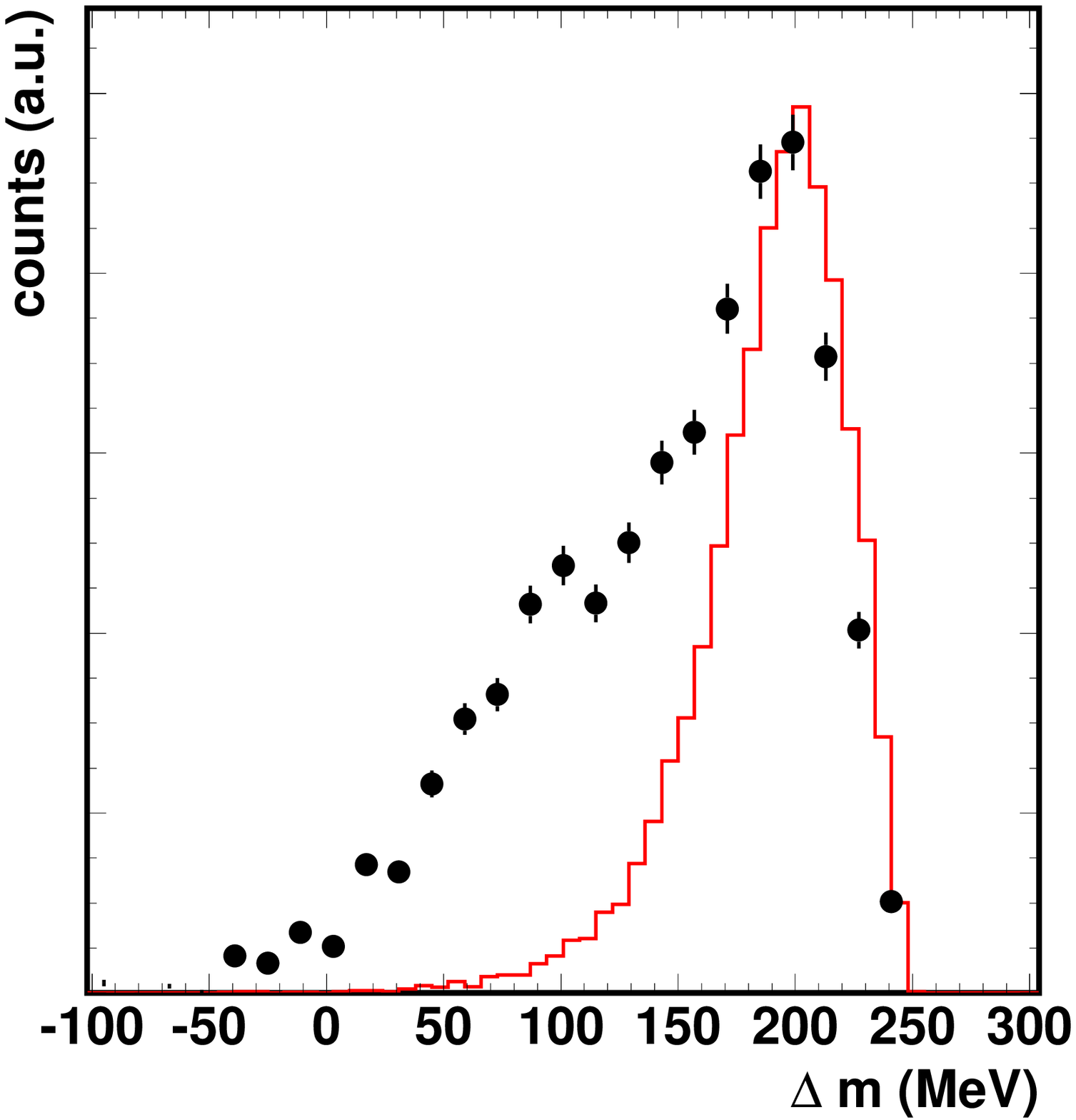}
}
\caption{Missing mass spectrum of $\pi^0\pi^+$ events for the background hypothesis
$p\pi^0$ ($E_{\gamma}$= 400 - 460 MeV). 
}
\label{fig:pi0p_mismas}       
\end{figure} 

The most efficient method for the identification of single $\pi^0$ photoproduction is
again the missing mass $\Delta m$ calculated under the assumption of a quasi-free 
reaction mechanism (analogous to eq. \ref{eq:mismas} with the sums running only 
over the two decay photons). The result of this analysis for the critical low energy
region is shown in fig. \ref{fig:pi0p_mismas}. The plot on the left hand side shows
the missing mass for the hypothesis of single $\pi^0$ production for events where 
only the detection of a neutral pion was required. It shows a clear peak around zero
corresponding to single $\pi^0$ production and the contribution from multiple pion
production at larger missing masses. The right hand side shows the spectrum under the
condition that an additional particle was detected which fulfills the
pion TOF-versus-E condition. The data are compared to a MC simulation of the
$\pi^o\pi^{\pm}$ final state. Some residual background from misidentified protons
is visible at missing masses lower than 150 MeV. For further analysis only events 
with a single $\pi^0$ missing mass above 150 MeV were accepted.  

Finally, the photoproduction of $\eta$ mesons was analyzed for the two decay modes
$\eta\rightarrow\gamma\gamma$ and $\eta\rightarrow\pi^0\pi^+\pi^-$. Since the cross
section for the $^{40}$Ca$(\gamma ,\eta)X$ reaction is known from previous 
experiments \cite{Roebig_96}, this reaction can be used to estimate systematic
effects in the detection of photons and charged pions and thus it helps to 
approximate the systematic uncertainties of the double pion production cross 
sections.  
The $\eta\rightarrow\gamma\gamma$ decay channel was identified with a
standard invariant mass analysis of photon pairs (see fig. \ref{fig:2ginv}, left hand
side). Events with invariant masses between 500 MeV and 600 MeV were selected.
This is the range where the invariant mass peak is practically free of background,
which has been investigated in detail for previous $\eta$ production experiments
with TAPS at MAMI at incident photon energies below 800 MeV
(\cite{Roebig_96,Krusche_95,Krusche_95a,Krusche_95b}. For the analysis of the  
$\eta\rightarrow\pi^0\pi^+\pi^-$ decay in a first step $\pi^0\pi^{\pm}$ pairs were
selected with the same cuts as for double pion production. In a second step 
a further hit fulfilling the TOF-versus-E condition for charged pions 
was required and finally the invariant mass of the three pions was constructed (see
fig. \ref{fig:2ginv}, right hand side). The spectrum is dominated by the invariant
mass peak of the $\eta$-meson, however there is also a small contribution from
$\pi^0\pi^{\pm}\pi^{\pm}$ final states not related to $\eta$ decays. Since the sum of
the mass of the three pions sets a low mass limit and the total available energy a
high mass limit of the distribution, the background populates similar invariant
masses as the $\eta$ decays and cannot be efficiently suppressed by a cut on invariant
mass. For further analysis all entries in the invariant mass spectrum have been
accepted, but the resulting cross section has been corrected for the obvious
background contribution at large invariant masses. For this correction the left hand
side of the peak was mirrored around the peak position (dashed histogram) and the
access on the right hand side above this mirror image was assigned to the background.
This procedure results most likely in a slight underestimation of the total 
background. 

\section{Determination of cross sections}
\label{sec:4}
The absolute normalization of the cross sections was obtained (see
ref. \cite{Krusche_99} for details) from the target density ((1.52$\pm$0.01)
g/cm$^2$), the photon flux, the decay branching ratio of
$\pi^0\rightarrow\gamma\gamma$, and the acceptance and detection efficiency of the 
TAPS detector. The latter was obtained from MC simulations with the GEANT3 package
\cite{Brun_86}. For the simulations the pion pairs have been generated according to
phase space distributions in quasifree kinematics taking into account the Fermi
motion of the nucleons. FSI was taken into account as in \cite{Messchendorp_02}
in a simple approximation, where pions scatter randomly off nucleons according to
their mean-free path. It was then checked, that missing mass-, kinetic energy-,
and angular distributions were in agreement with the simulations.

\begin{figure}[h]
\resizebox{0.45\textwidth}{!}{%
  \includegraphics{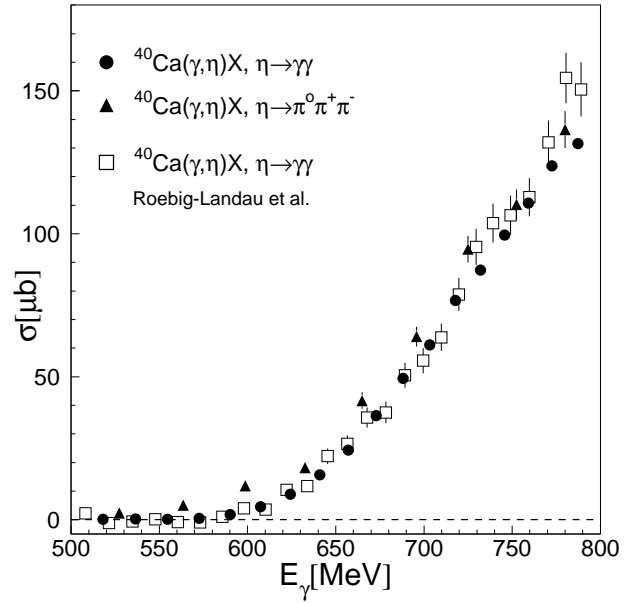}
}
\caption{Total cross sections for $\eta$ photoproduction off calcium. Compared are the
results for the $\eta\rightarrow\gamma\gamma$ and $\eta\rightarrow\pi^0\pi^+\pi^-$ 
decays to the results from a previous measurement \cite{Roebig_96}. 
}
\label{fig:eta}       
\end{figure}

For the extraction of the $\eta$ photoproduction cross section from the
$\eta\rightarrow\gamma\gamma$ decay the detection efficiency was obtained from
MC simulations as function of laboratory angle and energy of the mesons as in
\cite{Roebig_96}. This means that it becomes independent on assumptions about
energy or angular distributions of the mesons. This is not possible for the
$\eta\rightarrow\pi^0\pi^+\pi^-$ channel since the analysis used cuts on the 
reaction kinematic which depend on the incident photon energy. However, in this 
case the cross sections extracted from the $\eta\rightarrow\gamma\gamma$ channel 
can be used as input for the event generator. 
This is possible since the relatively
long-lived $\eta$ mesons decay always outside the nucleus so that the decay pions 
are not influenced by additional FSI effects.
The results for the total cross section of $\eta$ production are summarized in 
fig. \ref{fig:eta}.
The agreement between the previous data from \cite{Roebig_96} 
and the present results is quite good for both decay channels 
and can be used to estimate the systematic uncertainties. 

\section{Systematic uncertainties}
\label{sec:4b}
The systematic uncertainties for the two reaction channels and for different energy 
ranges arise from different sources, which are briefly discussed in this section.

\begin{figure}[h]
\resizebox{0.49\textwidth}{!}{%
  \includegraphics{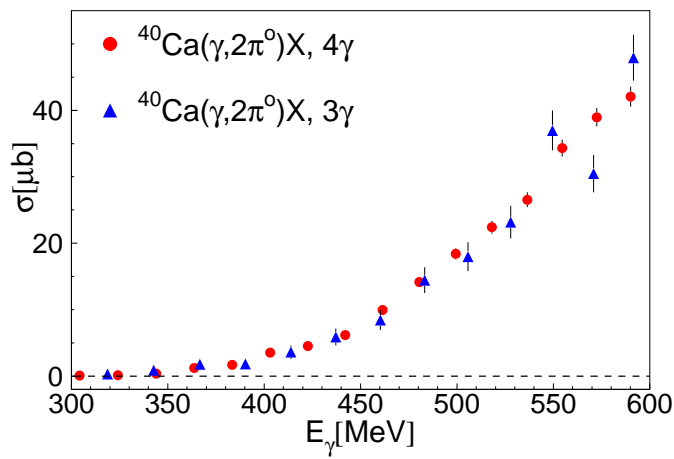}
}
\caption{Total cross sections for double $\pi^0$ production extracted from events with 
three \cite{Krusche_04} and four detected photons.
}
\label{fig:2pi0_mg}       
\end{figure}

We start with double $\pi^o$ photoproduction below the $\eta$ production threshold 
(i.e. for $E_{\gamma}\leq$ 550 MeV). The identification of photons in the TAPS 
detector is very clean (see e.g. ref. \cite{Krusche_95b}), and the suppression of 
background by the invariant mass analysis (see fig. \ref{fig:2pi0minv}) is efficient. 
As shown in fig. \ref{fig:2pi0mis} (left hand side), no background is visible in the 
missing mass spectrum for incident photon energies below 500 MeV, and the shape of the
missing mass spectrum is excellently reproduced by the Monte Carlo simulation.
The missing mass cut is therefore irrelevant in this energy region. Furthermore,  
in a previous experiment \cite{Krusche_04}, with an earlier stage of the TAPS detector,
the reaction was investigated below the $\eta$ threshold from an analysis of events
with three photons. In this case, only identification of one $\pi^0$ meson and an 
additional photon was required. Such an analysis was possible since 
the only physical background in this energy range comes from the reaction 
$\gamma N\rightarrow \gamma '\pi^0 N$. Compared to double $\pi^0$ production, the cross 
section of this background reaction is smaller by more than an order of magnitude 
\cite{Kotulla_02}, and the contribution to the three-photon sample was suppressed by 
another order of magnitude relative to the $\pi^0\pi^0$ channel via detection efficiency
(three-out-of-three versus three-out-of-four detected photons).
The result of this analysis for the total cross section
is compared in fig. \ref{fig:2pi0_mg} to the present result. The systematic
difference between the two excitation functions is only 3 \%, although the detector 
geometry, the cuts for the reaction identification, and the detection efficiencies 
were very different. We therefore estimate the systematic uncertainty of the double
$\pi^0$ data below the $\eta$-threshold with 5 \%.

At higher incident photon energies two effects increase the systematic uncertainty
for the double $\pi^0$ channel. Background from $\eta\rightarrow 3\pi^0$ decays,
where the decay photons from one pion escape detection, becomes important. At the same
time, the probability of re-scattering of the pions from the double $\pi^0$ production
in the nucleus increases, since the mean free path of pions decreases
strongly with increasing kinetic energy \cite{Krusche_04}. Together with the smearing
of the reaction kinematics due to Fermi motion, the re-scattering results in a
significant fraction of events which in the missing mass analysis overlap with
the $\eta$ background (see fig. \ref{fig:2pi0mis}). In the analysis, only events with 
missing masses between 100 and -100 MeV were accepted for double $\pi^0$ production.
The fraction of true 2$\pi^0$ events, which do not pass this cut, was determined from
the Monte Carlo simulations and corrected. The background from $\eta$ events which pass 
this cut can be subtracted, since the $\eta$ cross section is precisely known from the 
$\eta\rightarrow \gamma\gamma$ decay channel. This, however, involves a Monte Carlo 
simulation of $\eta\rightarrow 3\pi^0$ events. Both simulations contribute to the
systematic uncertainty. Among these two, the 2$\pi^0$ simulation is more critical 
since it involves model assumptions about the pion FSI. From a variation of
the missing mass cut we estimate the uncertainty at the 10 \% level for the highest
incident photon energies. Therefore a linearly rising systematic error 
(0 \% at 550 MeV, 10 \% at 800 MeV) is added in quadrature.

The systematic uncertainty for the mixed charge channel is larger since for the charged
pion we do not have the efficient invariant mass filter. A check of the simulated
detection efficiency for charged pions is obtained from the analysis of the 
$\eta\rightarrow \pi^0\pi^+\pi^-$ decay, where it enters squared. The agreement between
the cross sections from the two different $\eta$ decays (see fig. \ref{fig:eta}) limits 
the systematic uncertainty for detection of a $\pi^0\pi^{\pm}$ pair to the 15 \% range.
At low incident photon energies the main source for systematic uncertainties is
background due to misidentification of recoil protons from single $\pi^0$ 
production as charged pions (see sec. \ref{sec:4}), which is suppressed by a missing mass
cut (see fig. \ref{fig:pi0p_mismas}). Also from the comparison of the simulated missing 
mass spectrum to the measured distribution, a 15 \% uncertainty at the lowest incident
photon energies appears realistic. It should be noted that the fraction of true double 
pion events, which do not pass the 150 MeV cut, enters of course via the Monte Carlo
simulations into the detection efficiency. Systematic effects therefore arise only
from the correct representation of the double pion reaction in the simulation and from
residual proton background, which is not eliminated by the cut.
The relative importance of the proton background decreases with increasing incident
photon energy since the cross section ratio of single $\pi^0$ to $\pi^0\pi^{\pm}$
production drops by more than an order of magnitude between 450 and 600 MeV, although at the
same time the kinematic separation of the background gets worse due to pion re-scattering. 

The relative contribution of background from $\eta$-decays 
($\eta\rightarrow\pi^0\pi^+\pi^-$) is smaller by roughly an order of magnitude compared 
to the neutral channel. This is a combined effect of cross section ratios, $\eta$
decay branching ratios and misidentification probability ($\pi^0$ and one of the charged
pions must be detected while any pair out of three $\pi^0$'s is possible).
Therefore it was not attempted to suppress this background with further cuts,
but the simulated contribution was subtracted from the cross section.   
The related systematic uncertainty rises from threshold to high incident photon energies
to a few per cent and counteracts the dropping of the uncertainty related to the above 
discussed background from single $\pi^0$ production. In addition, at higher incident
photon energies a small contamination from $\pi^0\pi^{\pm}\pi^{\pm}$ events not originating
from $\eta$ decays may contribute (few per cent level, see fig. \ref{fig:2ginv}).
We have thus assumed a conservative total uncertainty of 15~\% for all incident photon 
energies.

Finally, an overall normalization uncertainty for the luminosity (target surface thickness
and photon flux) of 2 - 3 \% must be added.

\section{Results}
\label{sec:5}
\begin{figure}[h]
\resizebox{0.49\textwidth}{!}{%
  \includegraphics{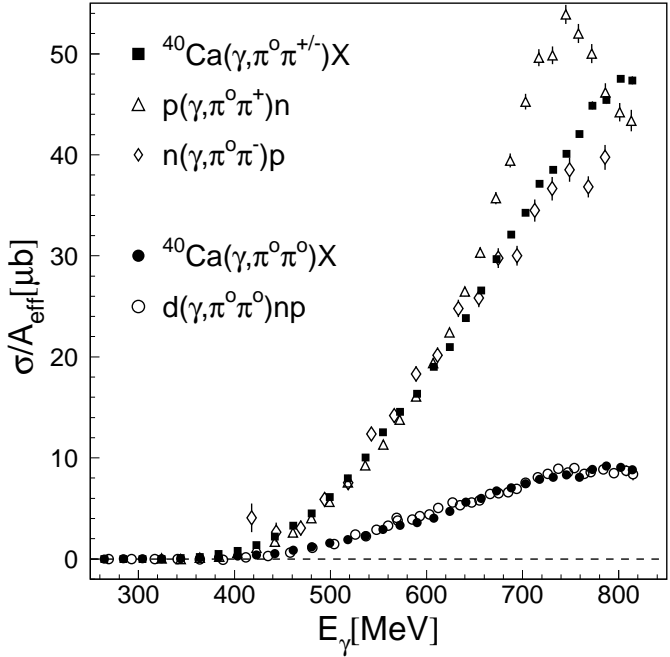}
}
\caption{Total cross sections for $\pi^0\pi^0$ and $\pi^0\pi^{\pm}$ photoproduction
off $^{40}$Ca compared to $d(\gamma ,\pi^0\pi^0)np$ \cite{Kleber_00} respectively 
to $p(\gamma ,\pi^0\pi^+)n$ \cite{Langgaertner_01} and $n(\gamma ,\pi^0\pi^-)p$
\cite{Zabrodin_97,Pedroni_05}. Cross sections are normalized to 
$A_{eff}$ = $A^{2/3}$ for calcium and $A_{eff}=A$ for $A=1,2$.
}
\label{fig:total_2pi}       
\end{figure}

The total cross sections for the two double pion production channels are compared in
fig. \ref{fig:total_2pi} to data off the nucleon. Two remarks must be made. 
The unpolarized double pion data off the nucleon measured with DAPHNE at MAMI
\cite{Braghieri_95,Zabrodin_97} was recently revised \cite{Ahrens_05}. 
The data shown for the $n(\gamma ,\pi^0\pi^-)p$ reaction are the result of this
re-analysis \cite{Pedroni_05}. The total cross section for $\pi^0\pi^{\pm}$ 
production off $^{40}$Ca from the TAPS experiment reported in \cite{Krusche_04}
is slightly higher than the present result (approximately 10 \% at 800 MeV). 
This discrepancy is due to the background from protons leaking into the pion
TOF-versus-E band, which was not efficiently enough suppressed in the 
earlier analysis. 

When properly scaled by the mass number dependence (as discussed in detail in
\cite{Krusche_04,Krusche_04a}) the total cross section for the $\pi^0\pi^0$ channel 
is in excellent agreement with the deuteron data over the full energy range,
indicating that only trivial effects like FSI influence the nuclear cross section.
The situation is less clear for the $\pi^0\pi^{\pm}$ reaction. The nuclear data
show a smooth behavior very similar to the $n(\gamma ,\pi^0\pi^-)p$ reaction, while
the $p(\gamma ,\pi^0\pi^+)n$ excitation function has a pronounced peak in the second
resonance region. This behavior might contribute to the still not understood complete
suppression of the second resonance bump in total photoabsorption data
\cite{Bianchi_94}. At least for this channel, a part of the effect might be due 
to a difference in proton and neutron cross section. However, since the 
$\pi^0\pi^+$ cross section is measured off the free proton and the $\pi^0\pi^-$ 
cross section off the neutron bound in the deuteron a more detailed investigation 
of possible nuclear effects in the deuteron is highly desirable. 

The main motivation for this experiment was the investigation of the $\pi -\pi$
invariant mass distributions in view of possible in-medium modifications of the
strength in the scalar, iso-scalar channel.   
For two reasons low incident photon energies are advantageous for this analysis.
First this avoids the additional background from triple pion decays of the 
$\eta$ meson, which would give rise to additional systematic uncertainties. 
Furthermore, at low incident photon energies the kinetic energies of the 
produced pions are small. This is advantageous because low energy pions, which 
have only a small chance to excite the nucleon to the $\Delta$ resonance, 
are much less affected by FSI \cite{Krusche_04}. On the other hand, the cross 
section drops strongly towards the production thresholds, so that some compromise 
is necessary.     

\begin{figure}[h]
\resizebox{0.49\textwidth}{!}{%
  \includegraphics{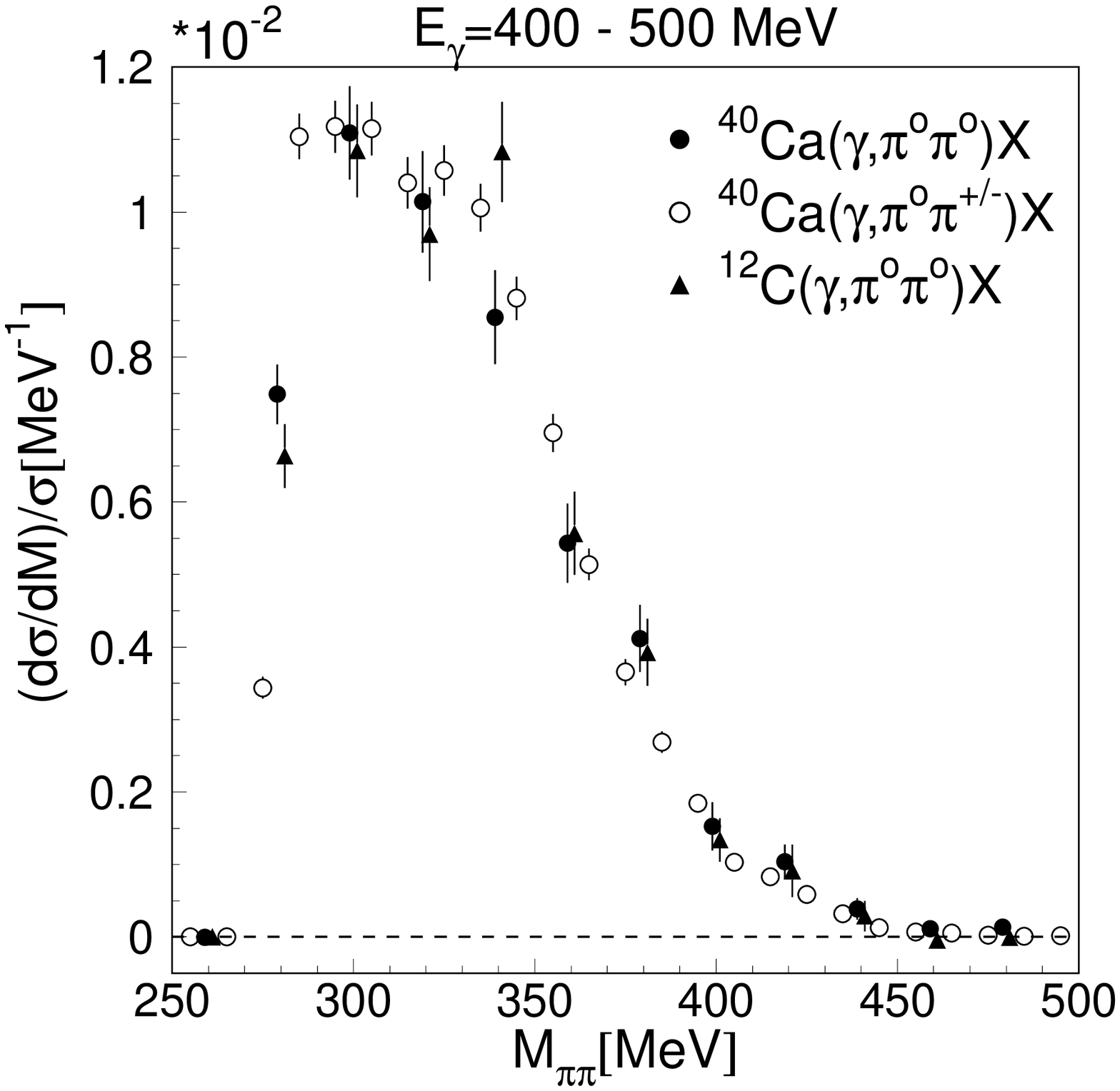}
  \includegraphics{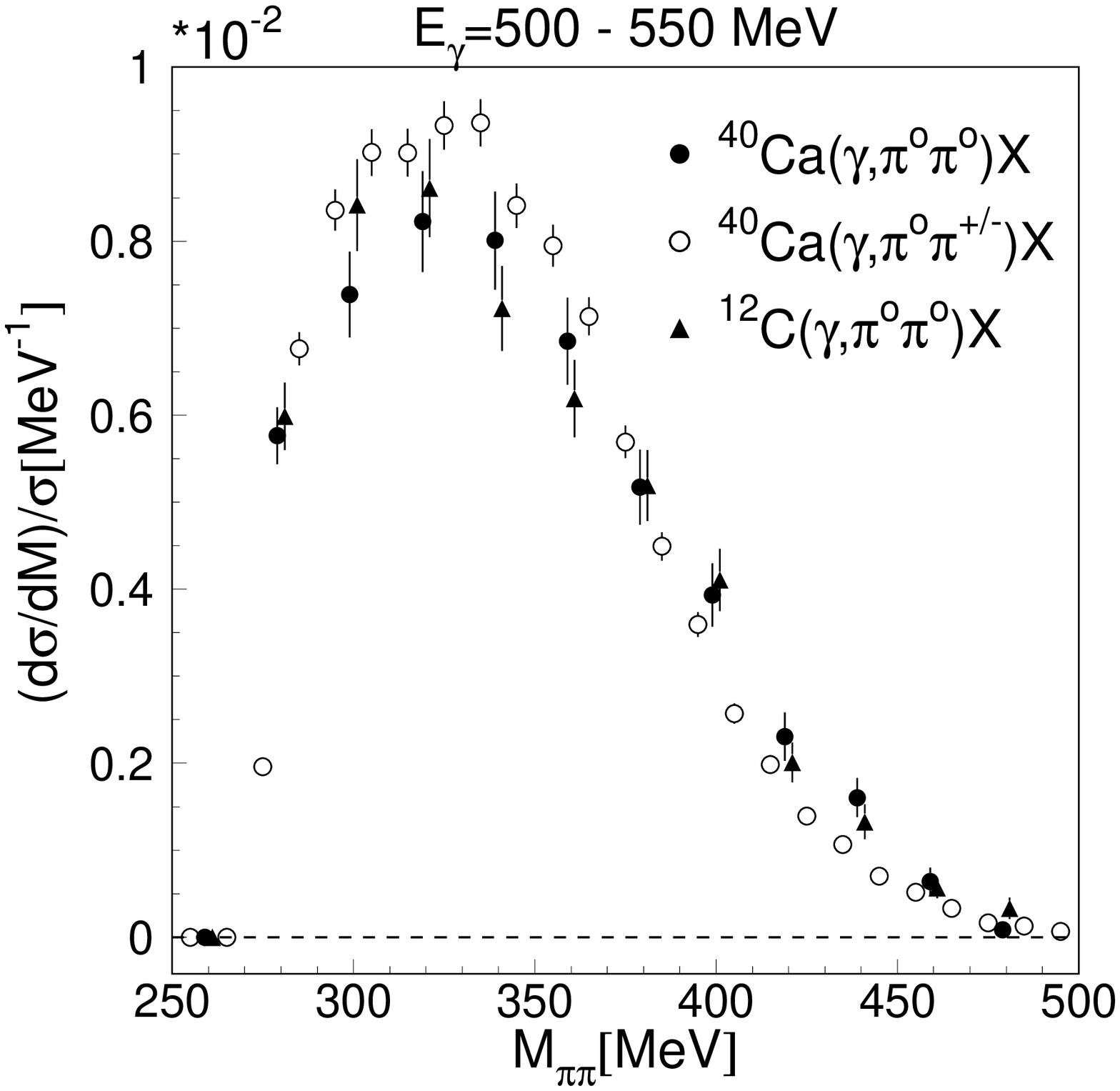}
}
\caption{Pion-pion invariant mass distributions for two different ranges of the
incident photon energy normalized to the total cross section. The distributions 
from $^{40}$Ca for $\pi^0\pi^0$ and $\pi^0\pi^{\pm}$ are compared to each other 
and to the $\pi^0\pi^0$ distributions from $^{12}C$ \cite{Messchendorp_02}.}
\label{fig:diff_1}       
\end{figure}

The measured distributions for two ranges of incident photon energies 
(400 MeV- 500 MeV and 500 MeV -550 MeV) are summarized in fig. \ref{fig:diff_1} and
compared to previous results for carbon \cite{Messchendorp_02}. The distributions 
for the two different charge channels and also for the two different nuclei are very 
similar. For $^{40}$Ca we do not observe such an important shift of the $\pi^0\pi^0$
invariant mass distributions towards small invariant masses as has been found for 
lead \cite{Messchendorp_02}, neither relative to the $\pi^0\pi^{\pm}$ channel 
for Ca nor in the comparison to $\pi^0\pi^0$ off carbon nuclei.

\begin{figure}[h]
\resizebox{0.49\textwidth}{!}{%
  \includegraphics{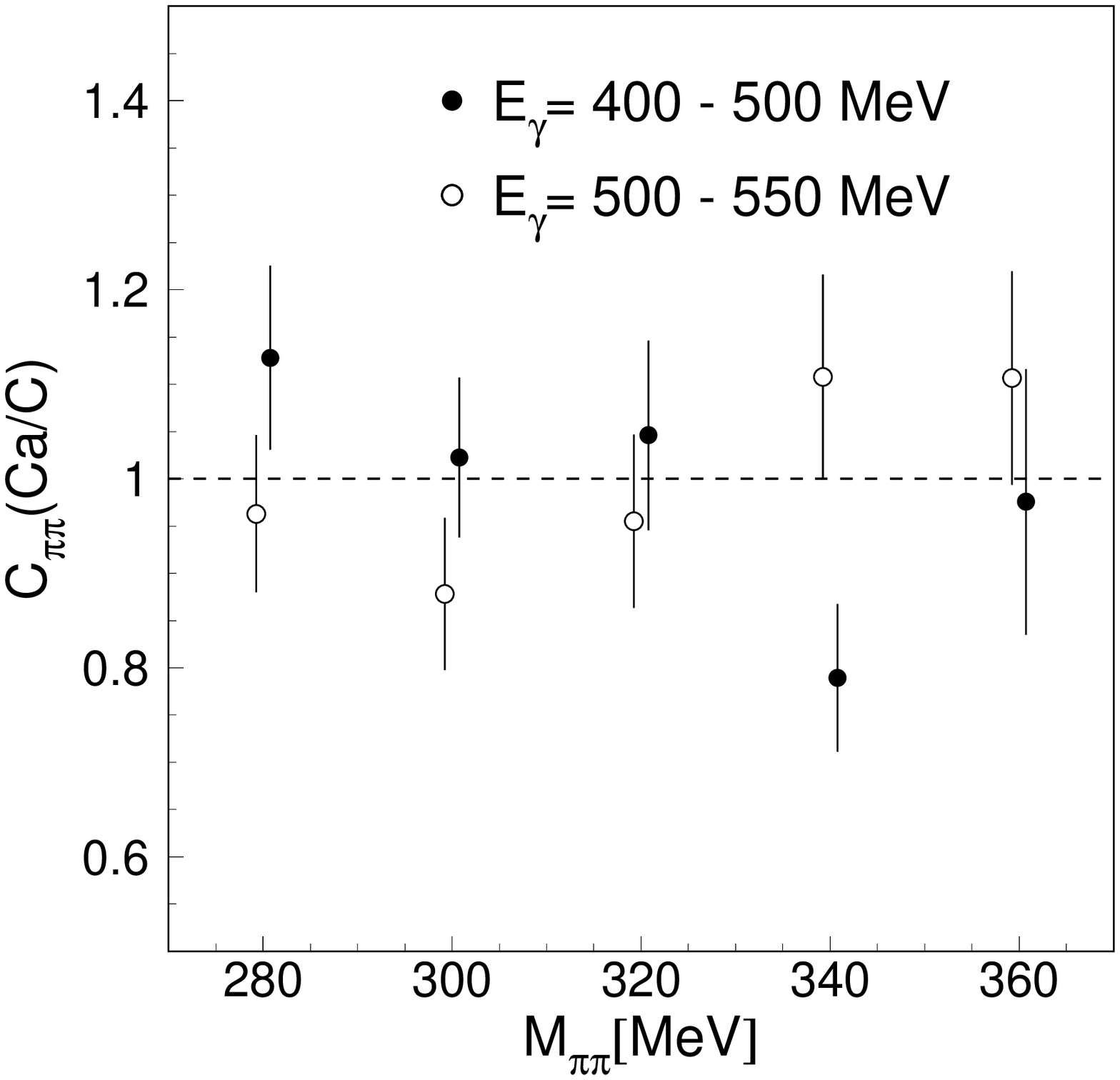}
  \includegraphics{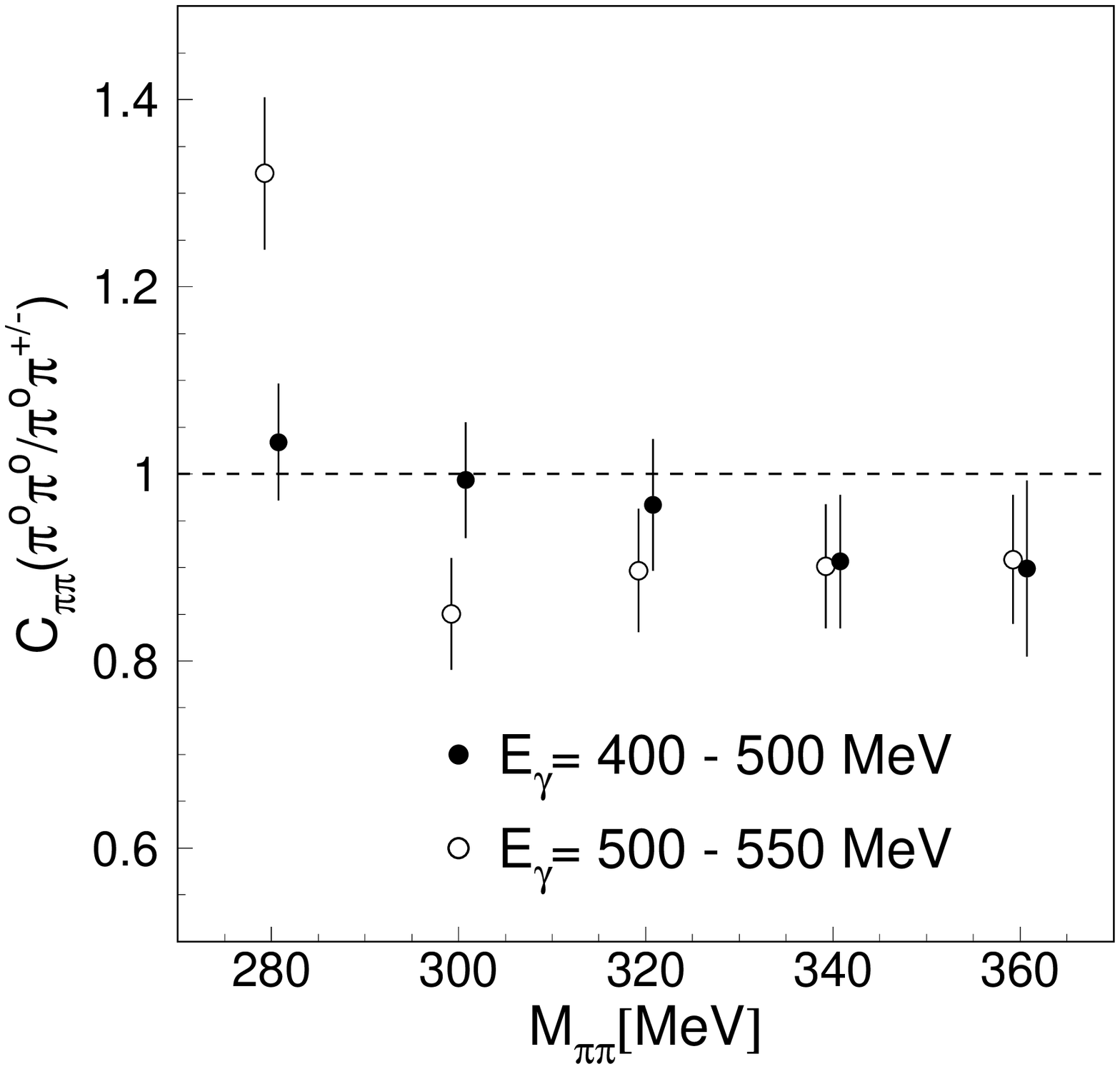}
}
\caption{Cross section ratios  
$C_{\pi\pi}(Ca/C)$ and $C_{\pi\pi}(\pi^0\pi^0/\pi^0\pi^{\pm})$ for low invariant
masses. 
}
\label{fig:ratio_1}       
\end{figure}

For a more detailed comparison the cross section ratios 
$C_{\pi\pi}(Ca/C)$ and $C_{\pi\pi}(\pi^0\pi^0/\pi^0\pi^{\pm})$
defined by:
\begin{eqnarray}
C_{\pi\pi}(Ca/C)\;\;\;\; & = &  
\frac{d\sigma_{Ca}(\pi^0\pi^0)}{\sigma_{Ca}(\pi^0\pi^0) dM} \left/ 
\frac{d\sigma_{C}(\pi^0\pi^0)}{\sigma_{C}(\pi^0\pi^0) dM}
\right.\\
C_{\pi\pi}(\pi^0\pi^0/\pi^0\pi^{\pm}) & = & 
\frac{d\sigma_{Ca}(\pi^0\pi^0)}{\sigma_{Ca}(\pi^0\pi^0) dM} \left/ 
\frac{d\sigma_{Ca}(\pi^0\pi^{\pm})}{\sigma_{Ca}(\pi^0\pi^{\pm}) dM}
\right.
\end{eqnarray}
where $\sigma_X(\pi^0\pi^c)$, $X$=carbon, calcium are the total cross sections 
and $d\sigma /dM$ the invariant mass distributions,
are shown in fig. \ref{fig:ratio_1}. In line with the previous results for carbon and
lead \cite{Messchendorp_02} some low mass enhancement of the $\pi^0\pi^0$ final state
versus the $\pi^0\pi^{\pm}$ final state is also visible for calcium. However, there is
no significant difference between the $\pi^0\pi^0$ distributions off carbon and off
calcium. This means that the size of possible in-medium effects is comparable for 
carbon and calcium, while significantly stronger effects had been found for lead 
\cite{Messchendorp_02}.    

\section{Comparison to results of the BUU-model}
\label{sec:6}
The quantitative discussion of the experimental results requires a detailed
understanding of the `trivial' in-medium effects like the smearing of cross sections
due to the momentum distributions of the bound nucleons, collisional broadening of 
nucleon resonances due to additional decay channels like $NN^{\star}\rightarrow NN$, 
Pauli-blocking of final states, and final state interactions of the pions.
These effects have been intensively studied in the framework of coupled channel 
transport models based on the semi-classical Boltzmann-Uehling-Uhlenbeck (BUU)
equation (see \cite{Effenberger_97,Teiss_97,Muehlich_03} for details of the model).
Recently M\"uhlich et al. \cite{Muehlich_04} and Buss et al. \cite{Buss_06} have 
performed detailed calculations of the nuclear double pion photoproduction reactions
in the framework of this model. In the latter work, special emphasis was laid on
the description of the scattering and absorption properties of low energy pions.

\begin{figure}[t]
\resizebox{0.49\textwidth}{!}{%
  \includegraphics{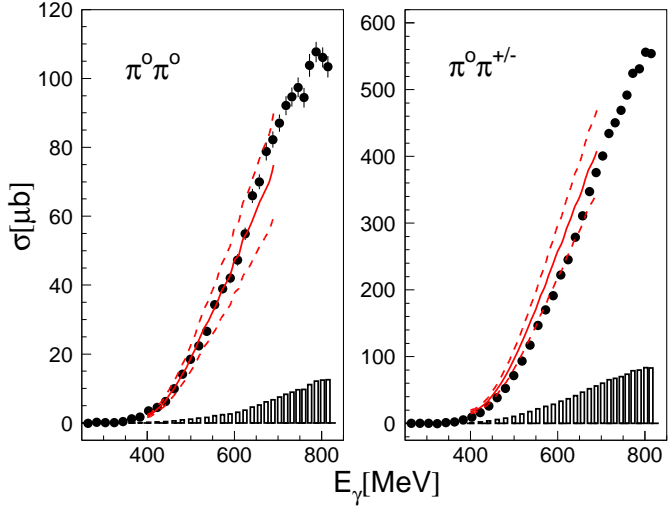}
}
\caption{Total cross sections compared to results of the BUU model \cite{Buss_06}.
The bars at the bottom represent the systematic uncertainty of the data 
(see sec. \ref{sec:4b}), the dashed lines represent the error band for the BUU 
calculation.}
\label{fig:total_theo}       
\end{figure}

\begin{figure}[t]
\resizebox{0.49\textwidth}{!}{%
  \includegraphics{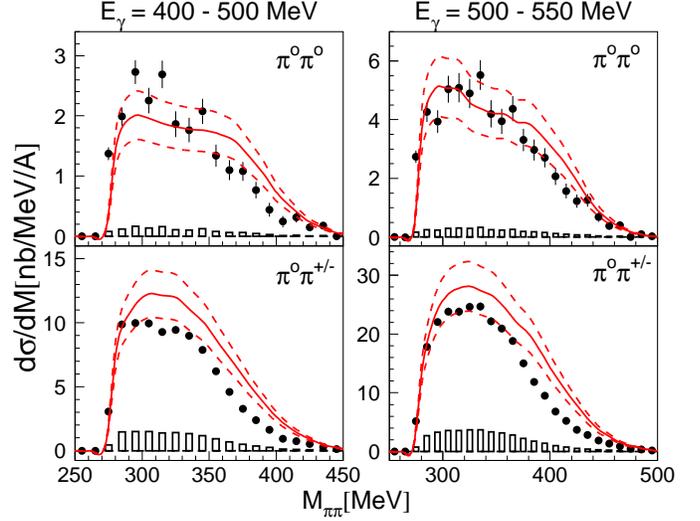}
}
\caption{Pion-pion invariant mass distributions compared to results of the BUU model
\cite{Buss_06}. The bars at the bottom represent the systematic uncertainty of the data 
(see sec. \ref{sec:4b}), the dashed lines represent the error band for the BUU 
calculation.
}
\label{fig:diff_theo}       
\end{figure}

Apart from the inherent assumptions and approximations made in semi-classical transport 
models, systematic uncertainties of the calculations arise from the uncertainty of
the cross sections of the elementary production reactions and the uncertainty of the 
cross sections involved in the description of pion FSI. Some indication for the correct
treatment of pion FSI comes from a comparison \cite{Buss_06a} of model calculations and 
experimental results for pion-induced double charge exchange reactions, which are very 
sensitive to details of pion FSI. It is thus assumed, that the major uncertainty is 
related to the experimentally determined elementary production cross sections of 
pion pairs off nucleons. Buss and co-workers \cite{Buss_06} estimate a systematic 20~\% 
uncertainty of their results for the $\pi^0\pi^0$ channel from a simultaneous variation 
of all input cross sections to their $\pm 1\sigma$ limits. This estimate is conservative
since no cancellation of systematic errors from different reaction channels is considered.
The largest uncertainty is related to the $\gamma n\rightarrow n\pi^0\pi^0$ channel.
The uncertainty estimated in \cite{Buss_06} for the $\pi^0\pi^{\pm}$ final state is much
smaller, however this was based only on the statistical uncertainties of the
input data and has neglected the systematic ones, which are also typically on the 
10 - 15~\% level. Therefore we assume also for this channel a 15 \% systematic uncertainty.
These model uncertainties and also the systematic uncertainties of the data are indicated
in figs. \ref{fig:total_theo},\ref{fig:diff_theo} as dashed lines, respectively bar charts.

The overall agreement between the model results and the data for the total cross
sections is good within the systematic uncertainties (see fig. \ref{fig:total_theo}), 
it is excellent for the $\pi^0\pi^0$ channel in the energy range of interest for the 
invariant mass distributions. Up to now, reliable BUU calculations are not available for
incident photon energies above 700 MeV, since in this region the elementary free 
nucleon cross sections have too large uncertainties for some channels.

An important result of this model calculations is, that even without any 
explicit in-medium modification of the $\pi^0\pi^0$ channel, the respective 
invariant mass distributions show a significant softening for heavy nuclei
\cite{Buss_06}.
In the model, this effect arises solely from FSI of the pions. Part of it comes
from the fact that the $\pi N$ absorption cross section increases strongly with pion
kinetic energy, so that pions with larger kinetic energies are more strongly depleted
via the $\pi N\rightarrow\Delta$, $\Delta N\rightarrow NN$ reaction path. 
But not only the absorptive part of the FSI is important, also important 
are re-scattering processes which tend to decrease the pion kinetic energy and 
thus the average pion - pion invariant masses. This effect is enhanced due to
charge exchange scattering, which mixes the contributions from the different charge
channels. Since the total cross section for $\pi^0\pi^{\pm}$ production is much 
larger than the $\pi^0\pi^0$ cross section, the latter receives significant side 
feeding from the mixed charge channel via $\pi^{\pm}N\rightarrow N\pi^0$ scattering,
which increases the fraction of re-scattered low-energy pions in this channel. 
In the same way, re-scattering of $\pi^+\pi^-$ contributes to the $\pi^0\pi^{\pm}$ 
channel.  
This result means, that the relative softening of the $\pi^0\pi^0$ invariant mass by 
itself is no evidence for an in-medium modification of the pion - pion interaction, 
but the argument becomes quantitative, requiring a comparison of the experimental 
results to calculations which account for the trivial FSI effects.

\begin{figure}[t]
\resizebox{0.49\textwidth}{!}{%
  \includegraphics{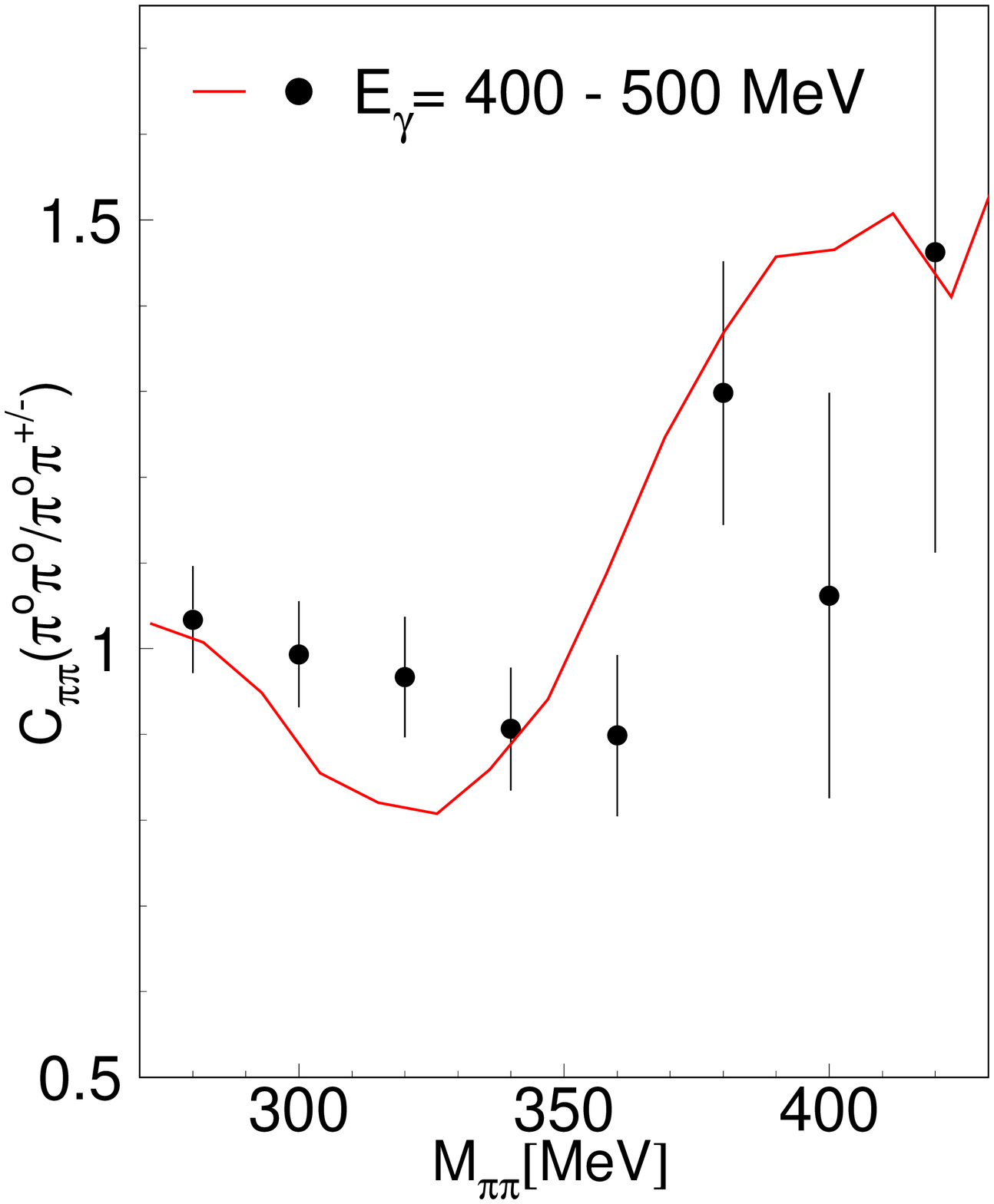}
  \includegraphics{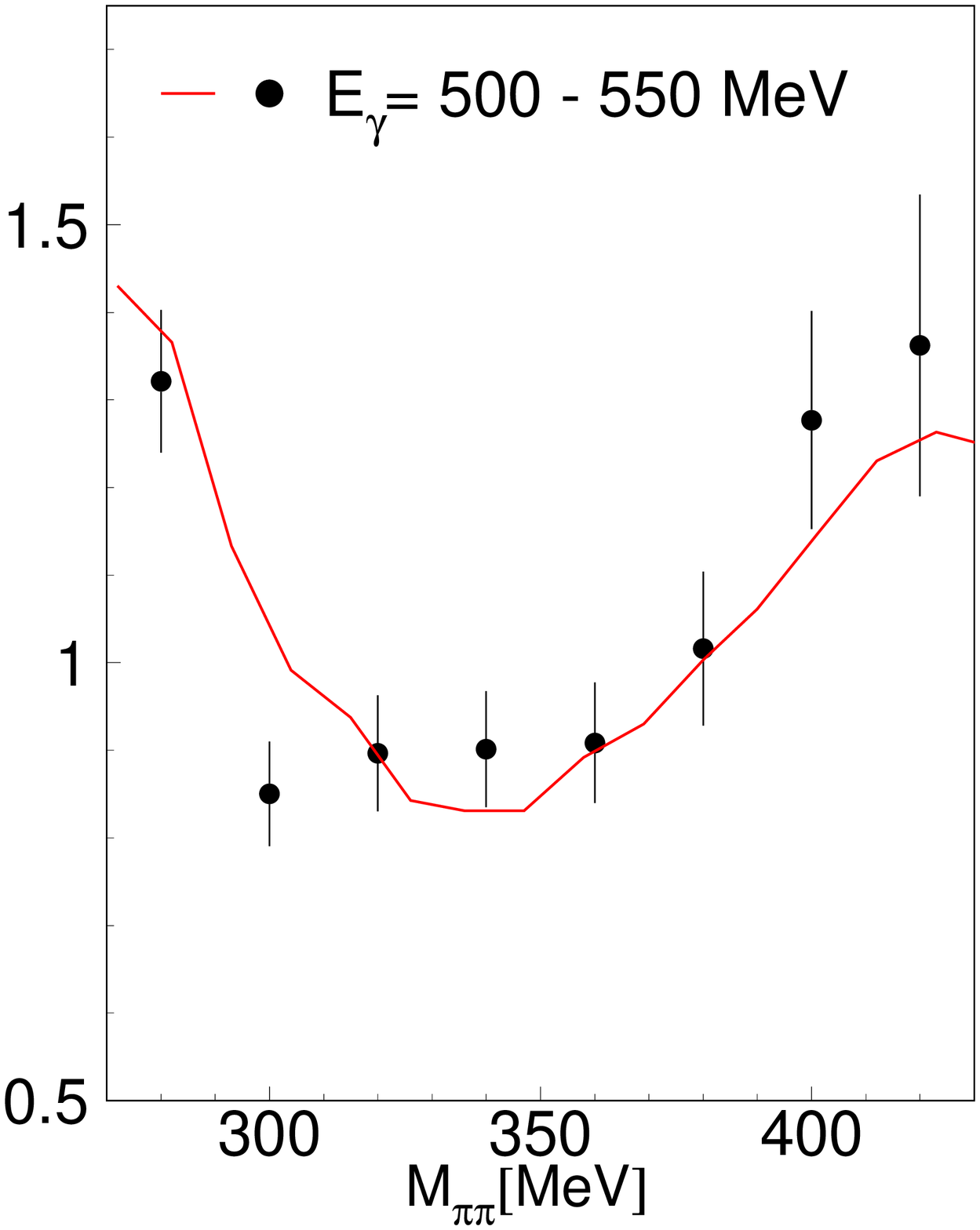}
}
\caption{Cross section ratio $C_{\pi\pi}(\pi^0\pi^0/\pi^0\pi^{\pm})$ compared to
the results of the BUU model \cite{Buss_06}. Symbols: data, curves BUU results
\cite{Buss_06}. Left hand side: incident photon energies
400 - 500 MeV, right hand side: incident photon energies 500 - 550 MeV.
}
\label{fig:ratio_buss}       
\end{figure}

The invariant mass distributions for the two energy ranges are compared to the BUU
results in fig. \ref{fig:diff_theo}. The general tendency of a softening of the
distributions is reproduced by the calculations as a consequence of FSI effects.
On a quantitative level, the data for the $\pi^0\pi^0$ invariant masses for the 
lowest incident photon energies seem to be slightly stronger downward shifted 
than predicted.  One should note that this effect cannot be completely explained by 
the systematic uncertainties of data and model results, since the uncertainties 
are not strongly invariant mass dependent but rather related to the absolute scale of 
the cross section. The cross section for the mixed charge channel is somewhat 
overestimated by the model, in particular for large invariant masses. This absolute 
normalization cancels in the $C_{\pi\pi}(\pi^0\pi^0/\pi^0\pi^{\pm})$ ratio, which is
compared to the BUU results in fig \ref{fig:ratio_buss}. Within the statistical
uncertainties, the agreement between data and model results is good. This means 
that the relative shapes of the $\pi^0\pi^0$ and $\pi^0\pi^{\pm}$ invariant mass 
distributions can be almost completely reproduced from FSI effects.

\section{Summary and conclusions}
\label{sec:7}

Total cross sections and pion-pion invariant mass distributions have been measured 
for $\pi^0\pi^0$ and $\pi^0\pi^{\pm}$ photoproduction off $^{40}$Ca nuclei.
When scaled by $A^{2/3}$ the total cross sections agree quite well with the average 
of the elementary cross sections off the proton and the neutron. The only exception 
is the mixed charge final state in the second resonance region, where the reaction 
off the free proton shows a more pronounced signal for the resonance bump.
 
The invariant mass spectra show a similar effect as already reported in
\cite{Messchendorp_02} for carbon and lead nuclei, namely a softening
of the $\pi^0\pi^0$ distributions relative to the $\pi^0\pi^{\pm}$ distributions.
The strength of the effect is comparable to carbon. 

The data have been compared to calculations in the framework of the BUU model
\cite{Buss_06}. A sizable part of the in-medium effects can be explained by the model
by final state interaction effects, which tend to shift re-scattered pions to smaller
kinetic energies. Only for the lowest incident photon energies a small additional
downward shift of the strength to small invariant masses for the $\pi^0\pi^0$ channel
may be visible.

\section{Acknowledgments}
We wish to acknowledge the excellent support of the accelerator group of MAMI,
as well as many other scientists and technicians of the Institut f\"ur
Kernphysik at the University of Mainz. We thank O. Buss, U. Mosel, and P. M\"uhlich 
for many valuable discussions and the communication of the BUU results. 
This work was supported by Schweizerischer Nationalfonds, Deutsche 
Forschungsgemeinschaft, and the UK Engineering and Physical Sciences
Research Council.

\end{document}